\documentclass[10pt, aps, prd, superscriptaddress, nofootinbib, showpacs, twocolumn]{revtex4-2}
\usepackage[english]{babel}
\usepackage{amsmath, amsfonts, amsthm, amssymb, mathrsfs}
\usepackage[all]{xy}
\usepackage{bm}
\usepackage{stmaryrd}
\usepackage{graphicx}
\usepackage[colorlinks]{hyperref}
\usepackage{color}
\usepackage{subcaption}
\usepackage{amscd}

\usepackage{pgfplots, grffile, tikz}
\pgfplotsset{compat=newest}
\usetikzlibrary{plotmarks, arrows.meta}
\usepgfplotslibrary{patchplots}

\usepackage{dsfont}

\DeclareMathOperator{\tr}{tr}
\DeclareMathOperator{\Tr}{Tr}
\renewcommand{\Re}{\mathop{\mathrm{Re}}\nolimits}

\newcommand{\calE}{\mathcal E}
\newcommand{\calK}{\mathcal K}
\newcommand{\calM}{\mathcal M}
\newcommand{\calR}{\mathcal R}

\newcommand{\bbB}{\mathbb B}
\newcommand{\bbR}{\mathbb R}
\newcommand{\bbW}{\mathbb W}
\newcommand{\bbZ}{\mathbb Z}

\newcommand{\frakL}{\mathfrak{L}}

\newcommand{\h}{\hat}

\begin{document}

\title{Pseudodifferential calculus in Schwinger--DeWitt formalism: UV and IR parts}

\author{A. O. Barvinsky}
\email{barvin@td.lpi.ru}
\affiliation{Theory Department, Lebedev Physics Institute, Leninsky Prospect 53, Moscow 119991, Russia}

\author{A. E. Kalugin}
\email{kalugin.ae@phystech.edu}
\affiliation{Theory Department, Lebedev Physics Institute, Leninsky Prospect 53, Moscow 119991, Russia}

\author{W. Wachowski}
\email{vladvakh@gmail.com}
\affiliation{Theory Department, Lebedev Physics Institute, Leninsky Prospect 53, Moscow 119991, Russia}

\begin{abstract}
We consider expansions for the kernels of operator functions of second-order minimal operators on a curved background. We show that the terms of these expansions originate in the ultraviolet or infrared regions. We propose a systematic approach to obtaining ultraviolet terms using term-by-term integration of the DeWitt expansion of the heat kernel. We discuss two methods for regularizing infrared divergences arising at intermediate computational steps---using analytic continuation and introducing a mass term---and the relationship between them.
\end{abstract}

\maketitle

\paragraph{Heat kernel method.}

A necessary step towards quantum gravity is the study of quantum field theories (QFTs) on a curved background. The basis for this approach is the remarkable DeWitt expansion for the heat kernel \cite{DeWitt1965, Gibbons, Birrell-Davies, Barvinsky1985}. Let us briefly recall its essence.

We consider the spacetime $\calM$ to be a $d$-dimensional Riemannian manifold with a metric $g_{ab}$. The theory is defined with a set of fields $\bm{\varphi} = \varphi^A$ which can be considered sections of a vector bundle over $\calM$ with the connection $\nabla_a$.\footnote{Spacetime indices are denoted by lowercase Latin letters $a, b, \ldots$. We omit the indices $A$ in the bundle, which can be of arbitrary nature and range over the tensor and spinor components of the fields, denoting matrices with these indices by a hat, for example, $\hat X=X^A_B$, $\tr\hat X=X^A_A$. We use dimensional regularization under which the dimension $d$ can formally take complex values. We work in the Euclidean version of the theory related to the physical Lorentzian signature spacetime by Wick rotation. $\nabla_a$ is an extension of the Levi--Civita connection. The Riemann tensor and curvature in the bundle are defined by the standard formulas: $[\nabla_a, \nabla_b] v^c = R^c{}_{dab} v^d$, $[\nabla_a, \nabla_b]\bm{\varphi} = \hat\calR_{ab}\bm{\varphi}$.}

If $\hat F(\nabla)$ is an elliptic, positive-definite differential operator acting on sections of the bundle, its heat kernel is the integral kernel of its operator exponent
\begin{equation} \label{HeatKernelDef}
\hat K_F(\tau| x,x') = e^{-\tau\hat F_x}\, \delta(x,x'),
\end{equation}
where $\delta(x,x')$ is the delta function. $\hat K_F(\tau| x,x')$ is a two-point matrix-valued function depending on two spacetime points $x,x'\in\calM$, as well as on an additional proper time parameter $\tau$.

In QFT in curved spacetime, the heat kernel is used to calculate the effective action. Namely, if $\hat F[\bm{\Phi}|\nabla]$ is the wave operator of a field theory (depending on nontrivial background fields $\bm{\Phi} = \Phi^A$), then the one-loop quantum effective action of this theory is given by
\begin{equation}\label{effective_action}
\varGamma_\text{1-loop}[\bm{\Phi}] = -\frac12 \Tr\ln\hat F = -\frac12 \int\limits_0^\infty \frac{d\tau}\tau \Tr e^{-\tau\hat F},
\end{equation}
where the functional trace is defined as
\begin{equation} \label{FuncTrDef}
\Tr e^{-\tau\hat F} = \int\limits_\calM d^dx\; \tr\hat K_F(\tau | x, x).
\end{equation}

For a second order minimal operator
\begin{equation} \label{minimal}
\hat F(\nabla) = -\Box + \hat P,
\end{equation}
where $\Box = g^{ab}\nabla_a\nabla_b$ is the covariant Laplacian and $\hat P(x)$ is a matrix depending on the point $x$ (a potential term), there is a remarkable asymptotic expansion~\cite{DeWitt1965} for the heat kernel $\hat K_F(\tau| x,x')$ as $\tau\to0$\footnote{A generalization of this expansion to the case of higher-order minimal operators $\hat F(\nabla) = (-\Box)^N + \hat P(\nabla)$ was constructed in \cite{Wach2, Wach3, BKW2024}, and a general algorithm for constructing similar expansions for so-called ``causal'' non-minimal operators was proposed in \cite{Barvinsky25}.}, which reads
\begin{equation} \label{HeatKernelExpansion}
\hat K_F(\tau | x,x') = \sum\limits_{k=0}^\infty B_{\frac{d}{2} - k}(\tau, \sigma) \cdot \hat a_k[F | x,x'],
\end{equation}
where the scalar function
\begin{equation} \label{InitialKernel}
B_\alpha(\tau, \sigma) = \frac{\tau^{-\alpha}}{(4\pi)^{d/2}} \exp\left(-\frac{\sigma}{2\tau}\right)
\end{equation}
depends neither on the bundle geometry nor on the potential term $\hat P$, and its dependence on the points $x$ and $x'$ is only implicit, through the Synge world function $\sigma(x,x')$, which is one half the square of the geodesic distance between $x$ and $x'$. On the other hand, the off-diagonal heat kernel coefficients $\hat a_m[F | x,x']$ (which are also known as HaMiDeW coefficients~\cite{Gibbons}) are matrix-valued two-point functions which do not depend on the proper time $\tau$, but contain information about the bundle geometry and the potential term $\hat P$.\footnote{It is always assumed that $x$ and $x'$ are sufficiently close to each other so that all quantities under consideration are well-defined. Also, note that the expansion \eqref{HeatKernelExpansion} is usually written with the quasiclassical prefactor of Pauli--van Vleck--Morette determinant which, in our notation, is absorbed into coefficients $\hat a_k[F| x,x']$.}

The coincidence limits of the HaMiDeW coefficients and their covariant derivatives can be found from the chain of recurrent equations for $\hat a_m[\,F\,|\,x,x']$ as some combinations of the background field curvatures $\mathfrak{R} \in \{R_a{}_{bcd}, \hat\calR_{ab}, \hat P\}$ and their derivatives \cite{DeWitt1965,Barvinsky1985}, which can be symbolically represented as
\begin{equation} \label{BackgroundDimensionEq}
\nabla^n \hat a_m[F] \Big|_{x=x'} \sim \sum\limits_{2k+l = 2m+n} \nabla^l \mathfrak{R}^k.
\end{equation}
Series \eqref{HeatKernelExpansion} substituted into \eqref{effective_action} yields a local expansion of the one-loop effective action $\varGamma_\text{1-loop}[\bm{\Phi}]$, a series in increasing powers of the background dimension.

\paragraph{Main idea.}

In many applications, there is a need for off-diagonal expansions similar to DeWitt's \eqref{HeatKernelExpansion} for kernels of operator functions of a more complicated nature than a simple operator exponential $\exp({-\tau\hat F})$. A typical example of such operator function is $\exp({-\tau\hat F})/\hat F^{\mu}$ which is useful for models with (degenerate and non-degenerate) causal non-minimal wave operators \cite{Barvinsky25, BarvinskyKalugin2024}.

At the same time, expansions for functional traces of the form $\Tr\big[Q(\hat F)\,\exp({-\tau\hat F})\big]$, where $Q(\hat F)$ is a polynomial, are often considered in mathematics~\cite{Seeley, Gilkey1975, Gilkey1995, GilkeyFegan}. Such objects possess remarkable functorial properties, however, due to the presence of functional trace \eqref{FuncTrDef} one is forced to consider the diagonal of corresponding integral kernels, which usually requires subtle refinement procedures using techniques such as $\zeta$-regularization and analytic continuation. Bypassing these issues could be possible should one employ an easier strategy of obtaining the diagonal kernels as coincidence limits of the corresponding off-diagonal objects.

The aforementioned off-diagonal expansion, essentially an analog of~\eqref{HeatKernelExpansion} for some operator function $f(\hat F)$, can be obtained from the DeWitt expansion itself. Notice, that the operator function $f(\h F)$ is obtained from the operator exponential $\exp({-\tau\hat F})$ using a linear integral transform $\mathfrak{L}_f$ with respect to the proper time parameter $\tau$:
\begin{equation} \label{frakLtransformEq}
f(\hat F) = \frakL_f\, e^{-\tau\hat F} = \int\limits_0^\infty d\tau\, f^*(\tau)\,e^{-\tau\hat F},
\end{equation}
where $f^*(\tau)$ is the inverse Laplace transform of $f(\lambda)$
\begin{equation} \label{InvLaplaceTransform}
f^*(\tau) = \int\limits_C \frac{d\lambda}{2\pi i}\; f(\lambda)\, e^{\tau\lambda}.
\end{equation}
Essentially,~\eqref{frakLtransformEq} can be understood as a direct Laplace transform from the variable $\tau$ to the operator variable $\hat F$. Rewriting this transform it terms of kernels, then substituting the DeWitt expansion \eqref{HeatKernelExpansion} into it, and finally calculating the proper time integrals \emph{term-by-term} we obtain the following generalization of the expansion \eqref{HeatKernelExpansion}:
\begin{equation} \label{general_kernel_series_rep}
f(\hat F_x)\, \delta(x,x') = \sum\limits_{k=0}^\infty \bbB_{\frac{d}{2}-k}\![f | \sigma] \cdot \hat a_k[F | x,x'],
\end{equation}
where $\hat a_k[ F\,|\,x,x']$ are exactly the same off-diagonal HaMiDeW coefficients as in the expansion \eqref{HeatKernelExpansion} and $\bbB_\alpha[f | \sigma]$ is an analytic function of the parameters $\sigma$ and $\alpha$, obtained from $B_\alpha(\tau, \sigma)$ \eqref{InitialKernel} by means of the aforementioned integral transform $\frakL_f$:
\begin{align}
&\bbB_\alpha[f | \sigma] = \frakL_f\, B_\alpha(\tau, \sigma) = \int\limits_0^\infty d\tau\, f^*(\tau)\, B_\alpha(\tau, \sigma) \nonumber \\
&= \frac{2}{(4\pi)^{d/2}} \int\limits_C \frac{d\lambda}{2\pi i}\, (-\lambda)^{\alpha-1}\, \calK_{\alpha-1}\!\left(-\frac{\sigma\lambda}{2}\right)\, f(\lambda), \label{frakLtransformEq2}
\end{align}
where
\begin{equation} \label{BC2Def}
\calK_\alpha(z) = \frac{1}{2} \int\limits_0^\infty dt\; t^{-\alpha-1} \exp\left(-t - \frac{z}{t}\right).
\end{equation}
is the Bessel--Clifford function of the second kind.

This procedure explicitly accounts for the separation of two different types of data into two different objects. In the expansion \eqref{general_kernel_series_rep}, all information about the bundle geometry and the operator $\hat F(\nabla)$ is still encoded in the HaMiDeW coefficients $\hat a_k[ F | x,x']$, while the functions $\bbB_\alpha\![f | \sigma]$, which we will call \emph{basis kernels}, do not depend either on the geometry, or on the specific form of the operator $\hat F(\nabla)$, but are determined exclusively by the function $f$.

The remarkable property that for any function $f$ the expansion \eqref{general_kernel_series_rep} includes the same HaMiDeW coefficients $\hat a_m[ F\,|\,x,x']$ is an off-diagonal generalization of the property that Gilkey and others call ``functoriality'' \cite{GilkeyFegan, Gilkey1975, Gilkey1995}. Therefore, we call it \emph{off-diagonal functoriality}. Off-diagonality is a new and key ingredient here, providing convenience and flexibility of the approach. This is due to the fact that, being essentially a form of point-separation regularization, it allows one to avoid dealing with the singularities that arise in the coincidence limit $x'\to x$, and to use the powerful apparatus of integral transforms more effectively.

Of course, this basic idea of term-by-term integration of the DeWitt expansion immediately encounters a simple objection: as is well-known, according to the Fubini--Tonelli theorem, one can change the order of summation/integration only if all the intermediate sums/integrals converge absolutely. However, the DeWitt expansion \eqref{HeatKernelExpansion} is not convergent, but is merely an asymptotic (i.e., divergent) series in the ultraviolet (UV) limit of small proper time $\tau\to 0$. Moreover, integrals appearing at the intermediate steps can also diverge at the infrared (IR) limit $\tau=\infty$. Despite this obvious difficulty, we argue that the idea of term-by-term integration is not as meaningless as it might seem at first glance.

\paragraph{Bessel--Clifford function example.}
To explain the implications of term-by-term integration of the DeWitt series, we consider an analytic function of a single complex variable, $\calK_\alpha(z)$ \eqref{BC2Def}, as a toy model. In this analogy, variable $t$ corresponds to the proper time $\tau$, the integrand $\exp(-t-z/t)$ is the heat kernel $\hat K_F(\tau | x, x')$, the integral over $t$ will be the integral transform $\frakL_f$ \eqref{frakLtransformEq}, and the function $\calK_\alpha(z)$ itself will serve as the kernel $f(\hat F)\, \delta(x, x')$.

We want to study the behavior of the function $\calK_\alpha(z)$ \eqref{BC2Def} in ``the coincidence limit'' $z\to0$. The naive logic described above amounts to expanding the factor $e^{-t}$ in the integrand into a power series in $t$, then swapping summation and integration, and calculating the integrals with help of the Euler integral $\Gamma(z) = \int\nolimits_0^\infty t^{z-1} e^{-t} dt$. This yields the following ``UV'' expansion:
\begin{align}
\calK_\alpha^\mathrm{UV}(z) &= \frac{1}{2} \sum\limits_{k=0}^\infty \frac{(-1)^k}{k!} \int\limits_0^\infty dt\; t^{k-\alpha-1} e^{-z/t} \nonumber \\
&= \frac{z^{-\alpha}}{2} \sum\limits_{k=0}^\infty \Gamma(\alpha-k) \frac{(-z)^k}{k!}.  \label{BCseries2}
\end{align}

On the other hand, we can perform a similar procedure in the opposite, ``IR'' region. Expansion of the factor $e^{-z/t}$ into a series and swapping summation and integration yields a completely different expansion:
\begin{align}
\calK_\alpha^\mathrm{IR}(z) &= \frac{1}{2} \sum\limits_{k=0}^\infty \frac{(-z)^k}{k!} \int\limits_0^\infty dt\; t^{-\alpha-k-1} e^{-t} \nonumber \\
&= \frac{1}{2} \sum\limits_{k=0}^\infty \Gamma(-\alpha-k) \frac{(-z)^k}{k!}. \label{BCseries1}
\end{align}

The fact that the ``UV'' \eqref{BCseries2} and ``IR'' \eqref{BCseries1} expansions do not coincide with each other $\calK_\alpha^\mathrm{UV}(z) \ne \calK_\alpha^\mathrm{IR}(z)$ is not surprising: it is easy to see that at least one of the integrals \eqref{BCseries2}-\eqref{BCseries1} diverges, so the trick with reversing the order of summation and integration, strictly speaking, does not work. However, what is truly remarkable and worthy of attention here is something entirely different: in fact, the correct asymptotics of the Bessel--Clifford function $\calK_\alpha(z)$ \eqref{BC2Def} is given by the sum of these ``UV'' and ``IR'' contributions\footnote{For the ``non-resonant'' case $\alpha\notin\bbZ$, the ``resonant'' case $\alpha\in\bbZ$ requires an additional limit be taken.}:
\begin{equation} \label{BCasymptotic}
\calK_\alpha(z) = \calK_\alpha^\mathrm{IR}(z) + \calK_\alpha^\mathrm{UV}(z).
\end{equation}
To see this, one finds the Mellin transform of the Bessel--Clifford function:
\begin{equation} \label{BCMeelinBarnes2}
\kappa_\alpha(s) = \int\limits_0^\infty \calK_\alpha(z) z^{s-1}dz= \frac{1}{2}\Gamma(s)\Gamma(s-\alpha).
\end{equation}
Then the inverse Mellin transform of~\eqref{BCMeelinBarnes2} gives the representation of the function $\calK_\nu(z)$ as the Mellin--Barnes integral
\begin{equation} \label{BCMeelinBarnes}
\calK_\alpha(z) = \int\limits_{w-i\infty}^{w+i\infty} \frac{ds}{2\pi i}\, z^{-s} \kappa_\alpha(s),
\end{equation}
where $w>\Re\alpha$.

Closure of the integration contour on the right reduces \eqref{BCMeelinBarnes} to the sum of the residues at the poles of $\kappa_\alpha(s)$. The sum of the residues at the poles of $\Gamma(s-\alpha)$ is exactly equal to the contribution $\calK_\alpha^\mathrm{UV}(z)$ \eqref{BCseries2}, while the sum of the residues at the poles of $\Gamma(s)$ is equal to the contribution $\calK_\alpha^\mathrm{IR}(z)$ \eqref{BCseries1}. Although these expansions were initially obtained from ``naive'' and ``illegal'' reasoning, they are not at all meaningless. In fact, they represent two contributions to the total asymptotics of $\calK_\alpha(z)$, coming from the ``UV'' region $t\to0$ and the ``IR'' region $t\to\infty$. Moreover, one can verify that this is not just a coincidence, but instead constitutes a universal phenomenon which holds true in all similar cases that come to mind.

\paragraph{Refined statement of the problem.}

We now return to our objective of calculating off-diagonal expansions for the integral kernel $f(\hat F)\, \delta(x, x')$, which is significantly more complex than a simple toy model we have considered above. One of the reasons for this jump in difficulty stems from the fact that UV and IR limits of the QFT in curved spacetime are fundamentally different. The source of this difference is that in this approach the physical spacetime is considered to be a Riemannian manifold. Therefore, locally, at small scales, it looks like Euclidean space $\bbR^d$, while its global, large-scale structure can be arbitrarily complex. A manifestation of this difference is that in the UV limit $\tau\to0$ one has the universal local DeWitt expansion \eqref{HeatKernelExpansion}, while in the limit $\tau\to\infty$, a similar universal expansion of the heat kernel simply does not exist. Its IR behavior depends strongly on the global structure of spacetime: its topology, properties of its boundary, boundary conditions, etc. In the physically interesting case of asymptotically flat spacetime, the corresponding $\tau\to\infty$ expansions were obtained in \cite{Barvinsky2002, Barvinsky03} within the framework of covariant perturbation theory \cite{CPTI, CPTII, CPTIIIa, CPTIII}.

Nevertheless, based on the analogy discussed above, we can assume that the expansion for the kernel of the operator function $f(\hat F)\, \delta(x, x')$ should be constructed (at least in the dimensionally regularized ``non-resonant'' case) from two independent parts---ultraviolet and infrared. The UV terms of the expansion can be obtained via term-by-term integration of the DeWitt expansion \eqref{HeatKernelExpansion}. And the IR terms could, in principle, be obtained from the asymptotic expansion of $\hat K_F(\tau | x, x')$ as $\tau\to\infty$, finding which is a significantly more subtle problem. Complete verification of our hypothesis requires finding these terms as well, at least for some simple manifolds like spheres $\mathbb{S}^d$ or asymptotically flat spaces. However, these considerations, being a matter of a future research, are beyond the scope of this Letter, in which we are primarily interested in the other, UV half of the expansion.

\paragraph{Regularization of IR divergences.}

However, even if we agree to limit our consideration to UV terms, we could still encounter IR divergent integrals, which arise at the intermediate steps of the calculation. Naturally, we need some method for their regularization. Two choices immediately come to mind.

Firstly, we can proceed as we did with the toy-model integrals \eqref{BCseries2}-\eqref{BCseries1}, i.e., employ a regularization via analytic continuation. In this case, we use the following simple rule: if the integral converges in a certain range of parameter values, then analytic continuation of the resulting expression beyond this range will yield the regularized value of the divergent integral. This method does not eliminate IR divergences completely, but it does allow us to isolate true physical divergences and deal with them in a controlled manner. Note, that despite the apparent hand-wavy nature of this approach, it can be rigorously justified in the Mellin transform framework, namely, in terms of the integration contour deformation.

The second idea is to employ a mass term $m^2$ as an IR regulator. The original operator $\hat F(\nabla)$ \eqref{minimal} can itself be massive, i.e., initially contain such a term; and should $\hat F(\nabla)$ be massless, one can try to introduce it artificially. In both cases, the presence of the prefactor $e^{-\tau m^2}$ in the DeWitt expansion will lead to convergence of all integrals at the IR limit $\tau=\infty$. This procedure will give us a new expansion:
\begin{equation} \label{MassiveExpansion}
f(\hat F_x + m^2)\, \delta(x,x') = \sum\limits_{k=0}^\infty \bbW_{\frac{d}{2}-k}\![f | \sigma, m^2] \cdot \hat a_k[F | x,x'].
\end{equation}
It can be obtained from the expansion \eqref{general_kernel_series_rep} by replacing the basis kernels $\bbB_\alpha[f | \sigma]$ with some new  well-defined objects:
\begin{align}
&\bbW_\alpha\!\big[f \big| \sigma, m^2\big] = \frakL_f\; W_\alpha(\tau, \sigma, m^2), \label{BHtransform} \\
&W_\alpha(\tau, \sigma, m^2) = \frac{\tau^{-\alpha}}{(4\pi)^{d/2}} \exp\left(-\frac{\sigma}{2\tau} - m^2 \tau\right), \label{IRregKernel}
\end{align}
which we will call \emph{complete massive kernels}.

The relationship between the two regularization approaches is as follows: the original basis kernel expansion \eqref{general_kernel_series_rep} with the regularization procedure via analytic continuation captures the terms coming from the UV region exactly and does not include the IR terms of the total expansion. When we use the expansion with complete massive kernels \eqref{BHtransform}, the massive term enter ``non-perturbatively.'' In this case, the IR asymptotic behavior of the heat kernel associated with the factor $e^{-\tau m^2}$ is partially taken into account. Therefore, some additional IR terms also appear in this expansion.

The treatment of these additional IR terms should be determined by the physical meaning of the massive term $m^2$. If the original wave operator of the theory $\hat F(\nabla)$ was massive, and if we do not want to lose the IR terms associated with the exponential decay of $e^{-\tau m^2}$, then we are simply obliged to use the expansion with complete massive kernels \eqref{BHtransform}, rather than with basis kernels \eqref{general_kernel_series_rep}.

A completely different situation occurs if the operator $\hat F(\nabla)$ was initially massless, and mass was introduced only in an attempt to regularize IR divergent integrals. Then the generated IR terms (just like the corresponding nonlocal terms in the effective action) actually carry no physical meaning, and should be regarded as an artifact of the method used. Therefore, in massless theories we are restricted to using the expansion with basis kernels \eqref{general_kernel_series_rep} while employing regularization via analytic continuation, which in this case is a more correct procedure. However, if one wishes to take the physical IR terms into account, one has to devote oneself to a much more careful investigation of the IR behavior of the heat kernel.

In the Appendix we look at a couple of simple examples that support these claims. In upcoming papers \cite{BKW25a, BKW25b}, we consider the issues raised in this Letter in much greater detail. In particular, we calculate the basis $\bbB_\alpha$ \eqref{frakLtransformEq2} and complete $\bbW_\alpha$ \eqref{BHtransform} kernels for operator functions of a more complex form, such as $e^{-\tau\hat F}/(F^\mu + \lambda)$. It turns out that these kernels can always be represented as $N$-fold Mellin--Barnes integrals (where $(N+1)$ is the number of dimensional parameters in the problem). This universal representation determines the greater practical efficiency and convenience of the developed technique.

\paragraph*{Acknowledgments:}
The work of AOB and AEK was supported by the grant from the ``BASIS'' Foundation for the Advancement of Theoretical Physics and Mathematics.

\appendix
\section{Examples}

\paragraph{Basis kernels for $\hat F^{-\mu}$ and $\exp(-\tau\hat F^\nu)$.}

The operator complex power is defined using a relation known in QFT as the ``Schwinger representation''
\begin{equation} \label{CompPowDef}
\hat F^{-\mu} = \frac{1}{\Gamma(\mu)} \int\limits_0^\infty d\tau\; \tau^{\mu-1}\, e^{-\tau\hat F},
\end{equation}
where $\mu\ne 0, -1, -2, \ldots$. The inverse transform to~\eqref{CompPowDef} reads:
\begin{equation} \label{CompPowInw}
e^{-\tau\hat F} = \int\limits_{w-i\infty}^{w+i\infty} \frac{d\mu}{2\pi i}\, \tau^{-\mu}\, \Gamma(\mu) \,\hat F^{-\mu},
\end{equation}
where $w>0$. Acting with operator functions involved in these relations upon $\delta(x,x')$, we find that the heat kernel $\hat K_F(\tau | x, x')$ and the Green function $\hat G_{F^\mu}(x, x')$ are related to each other by the direct and inverse Mellin transforms.

Therefore, the basis kernel $\bbB_\alpha$ for complex power $\hat F^{-\mu}$ is given simply by the Mellin transform of the function $B_\alpha(\tau, \sigma)$ \eqref{InitialKernel}:
\begin{align}
\bbB_\alpha\!\big[F^{-\mu} \big| \sigma\big] &= \frac{1}{(4\pi)^{d/2}\Gamma(\mu)} \int\limits_0^\infty d\tau\; \tau^{\mu-\alpha-1} e^{-\sigma/2\tau} \nonumber\\
&= \frac{\Gamma\left(\alpha-\mu\right)}{(4\pi)^{d/2}\Gamma(\mu)} \left(\frac{\sigma}{2}\right)^{\mu-\alpha}. \label{CompPowFunctions}
\end{align}
The integral above diverges at the IR limit $\tau=\infty$ for $\Re(\alpha-\mu) < 0$. Borrowing terminology from renormalization theory, it is appropriate to call this parameter region ``irrelevant,'' and the complementary region $\Re(\alpha-\mu) > 0$ ``relevant.'' Employing analytic continuation regularization we postulate that the basis kernels $\bbB_\alpha\!\big[F^{-\mu} \big| \sigma\big]$ are given by the expression \eqref{CompPowFunctions} in the irrelevant region, where the integral diverges. Note that since $\alpha = d/2 - k$, there is only a finite number of ``relevant'' terms in this expansion and an infinite number of ``irrelevant'' terms.

As noted above, analytic continuation does not eliminate IR divergences completely: \eqref{CompPowFunctions} still tends to infinity at poles of $\Gamma(\alpha-\mu)$. Therefore infinitely many terms of the expansion \eqref{general_kernel_series_rep} for $\hat G_{F^\mu}(x, x')$ will diverge in the physical case of even spacetime dimension $d$ and integer $\mu$ (or odd dimension $d$ and half-integer $\mu$). Unlike the divergence of the integral in \eqref{CompPowFunctions} in the irrelevant region, we interpret these poles as true, physical IR divergences.

We can now obtain the basis kernel $\bbB_\alpha$ for $\exp(-\tau\hat F^\nu)$ by applying the inverse Mellin transform to the basis kernel \eqref{CompPowFunctions}:
\begin{align}
\bbB_\alpha\!\big[e^{-\tau F^\nu} \big| \sigma\big] &= \int\limits_C \frac{d\rho}{2\pi i}\, \tau^{-\rho}\,\Gamma(\rho)\;\bbB_\alpha\!\left[F^{-\rho\nu} | \sigma\right] \nonumber \\
&= \frac{\tau^{-\frac{\alpha}{\nu}}}{(4\pi)^{d/2}}\, \calE_{\nu,\alpha}\!\left(-\frac{\sigma}{2\tau^{1/\nu}}\right). \label{PowHeatKernel}
\end{align}
Here we reduced the result to \emph{``generalized exponential functions'' (GEFs)} $\calE_{\nu,\alpha}(z)$ introduced in \cite{Wach2}. These functions can be defined in terms of the Mellin--Barnes integral:
\begin{align}
\calE_{\nu,\alpha}(z) &= \int\limits_C \frac{ds}{2\pi i}\,(-z)^{-s}\, \varepsilon_{\nu,\alpha}(s) \nonumber \\
&= \frac{1}{\nu} \sum\limits_{m=0}^\infty \frac{\Gamma\left(\frac{\alpha+m}{\nu}\right)}{\Gamma(\alpha+m)} \frac{z^m}{m!}, \label{InvMellinCalE} \\
\varepsilon_{\nu,\alpha}(s) &= \int\limits_0^\infty dz\, z^{s-1}\,\calE_{\nu, \alpha}(-z) = \frac{\Gamma(s)\Gamma\left(\frac{\alpha-s}{\nu}\right)}{\nu\Gamma(\alpha-s)}. \label{MellinCalE}
\end{align}
Their properties were studied in detail in~\cite{Wach2}, where we refer the interested reader to. For now, it is important for us that expressions \eqref{PowHeatKernel}-\eqref{InvMellinCalE} are well defined and no longer contain any divergences for positive integer $\nu$.

Since $\calE_{\nu,\alpha}(0) = \Gamma(\alpha/\nu)/\nu\Gamma(\alpha)$ and the basis kernels \eqref{PowHeatKernel} are well defined in the coincidence limit $\sigma\to0$, expansion \eqref{general_kernel_series_rep} for $\exp(-\tau\hat F^\nu)$ can be used to derive known ``diagonal'' relations. Namely, if the Seeley--Gilkey coefficients $\hat E_k[H | x]$ for the differential operator $\hat H(\nabla)$ of order $2\nu$ are defined in accordance with a diagonal expansion
\begin{equation}
\hat K_H(\tau | x, x) = \tau^{-d/2\nu} \sum\limits_{l=0}^\infty \tau^{k/\nu}\, \hat E_{2k}[H| x],
\end{equation}
then, taking the coincidence limit $\sigma\to0$ of the expansion \eqref{general_kernel_series_rep}, we restore the well-known Fegan--Gilkey formula~\cite{GilkeyFegan}:
\begin{equation} \label{FeganGilkey}
\hat E_k[F^\nu | x] = \frac{\Gamma\left(\frac{d - k}{2\nu}\right)}{\nu\Gamma\left(\frac{d - k}{2}\right)}\, \hat E_k[F | x].
\end{equation}

\paragraph{The complete massive kernels for $\hat F^{-\mu}$.}

In exactly the same way, applying the Mellin transform to the function $W_\alpha(\tau, \sigma, m^2)$ \eqref{IRregKernel} yields the complete massive kernel for $\hat F^{-\mu}$:
\begin{align}
&\bbW_\alpha\!\big[F^{-\mu} \big| \sigma, m^2\big]= \frac{1}{\Gamma(\mu)} \int\limits_0^\infty d\tau\; \tau^{\mu-1}\, W_\alpha(\tau, \sigma, m^2) \nonumber \\
&\qquad= \frac{2m^{2(\alpha-\mu)}}{(4\pi)^{d/2} \Gamma(\mu)}\; \calK_{\alpha-\mu}\!\left(\sigma m^2/2 \right). \label{GreenIRreg}
\end{align}
This expression is a massive analogue of \eqref{CompPowFunctions}, but the integral in it converges at the IR limit $\tau=\infty$, so it does not have any poles.

To obtain either the coincidence or the massless limit of the obtained expression, we substitute the leading term of the asymptotics \eqref{BCasymptotic} into it. Note that this yields different answers for the relevant $\Re(\alpha-\mu) > 0$ and irrelevant $\Re(\alpha-\mu)<0$ parameter regions. Indeed, in the relevant region, the leading term comes from the ``UV'' part of the Bessel--Clifford function expansion $\calK_\alpha^\mathrm{UM}(z)$ \eqref{BCseries2}. In this case, the dependence on the mass term $m^2$ completely disappears, and we reproduce the previously obtained answer for massless basis kernels \eqref{CompPowFunctions} (divergent in the limit $\sigma\to0$):
\begin{equation} \label{LimitM2to0Conv}
\bbW_\alpha\!\big[ F^{-\mu} \big| \sigma, m^2\big] \xrightarrow[m^2\to0]{\mathrm{relevant}} \bbB_\alpha\!\big[ F^{-\mu} \big| \sigma\big].
\end{equation}

Conversely, in the irrelevant parameter region, the leading term comes from the ``IR'' part of the Bessel-Clifford function expansion $\calK_\alpha^\mathrm{IR}(z)$ \eqref{BCseries1}. This time, the dependence on $\sigma$ completely disappears, and we obtain a completely different expression (diverging in the massless limit $m^2\to0$):
\begin{equation} \label{DivergenceRegionLimit}
\bbW_\alpha\!\big[ F^{-\mu} \big| \sigma, m^2\big] \xrightarrow[\sigma\to0]{\mathrm{irrelevant}} \frac{\Gamma\left(\mu-\alpha\right)}{(4\pi)^{d/2}\Gamma(\mu)} m^{2(\alpha-\mu)}.
\end{equation}
When calculating the quantum effective action in the background field method, an expansion in powers of the background dimension arises just from this expression.

\paragraph{Resummation of series.}

To illustrate the relation between the massive and analytic regularizations, note that in the former one can introduce the massive term $m^2$ in two different ways: either by ``non-perturbatively'' including $e^{-\tau m^2}$ into the prefactor of \eqref{HeatKernelExpansion}, or by considering $m^2$ as a part of the operator $\hat F$ thus ``perturbatively'' including mass into the HaMiDeW coefficients. Since for the heat kernel $\hat K_{F+m^2}(\tau | x, x')$ both ways are equivalent, this leads to the following transformation rule for HaMiDeW coefficients:
\begin{equation} \label{hatATransform1}
\hat a_k[F+m^2 | x, x'] = \sum\limits_{j=0}^k \frac{(-m^2)^{k-j}}{(k-j)!} \hat a_j[F | x, x'].
\end{equation}

However, since, as we discussed above, term-by-term integration only accurately reflects the UV behavior, after applying it, these two approaches will lead to different expansions. For example, for the massive Green function $\hat G_{(F+m^2)^\mu}(x,x')$, we obtain two expansions:
\begin{align}
\hat G_\text{pert} &= \sum\limits_{k=0}^\infty \bbB_{\frac{d}{2} - k}\!\big[F^{-\mu} \big| \sigma\big] \cdot \hat a_k[F + m^2], \label{PerturbExpansion} \\
\hat G_\text{non-pert} &= \sum\limits_{k=0}^\infty \bbW_{\frac{d}{2} - k}\!\big[F^{-\mu} \big| \sigma, m^2\big] \cdot \hat a_k[F]. \label{MassiveGreenExpansion}
\end{align}
Using the transformation rule \eqref{hatATransform1} we can compare them.

According to \eqref{GreenIRreg}, $\hat G_\text{non-pert}$ includes the Bessel--Clifford functions $\calK_\alpha(z)$. In accordance with the above decomposition of their asymptotics into the ``UV'' and ``IR'' parts \eqref{BCasymptotic}, we can do the same with the expansion \eqref{MassiveGreenExpansion}: $\hat G_\text{non-pert} = \hat G_\text{UV} + \hat G_\text{IR}$. Then using the transformation rule~\eqref{hatATransform1}, it is easy to verify, that the ultraviolet part of the expansion can be exactly resummed into $\hat G_\mathrm{pert}$ \eqref{PerturbExpansion}:
\begin{align}
\hat G_\mathrm{UV} &= \frac{2m^{d-2\mu}}{(4\pi)^{d/2}\Gamma(\mu)} \sum\limits_{k=0}^\infty m^{-2k} \calK_{\tfrac{d}{2}-\mu-k}^\mathrm{UV}\!\left(\frac{\sigma m^2}{2}\right) \cdot \hat a_k[F] \nonumber \\
&= \frac{(\sigma/2)^{\mu-d/2}}{(4\pi)^{d/2}\Gamma(\mu)} \sum\limits_{k=0}^\infty \sum\limits_{l=0}^\infty \frac{(-m^2)^l}{l!} \left(\frac{\sigma}{2}\right)^{k+l} \nonumber \\
&\times\Gamma\left(\tfrac{d}{2}-\mu-k-l\right) \cdot \hat a_k[F] = \hat G_\mathrm{pert}. \label{UVterms}
\end{align}
At the same time, the infrared part
\begin{align}
\hat G_\mathrm{IR} &= \frac{2m^{d-2\mu}}{(4\pi)^{d/2}\Gamma(\mu)} \sum\limits_{k=0}^\infty m^{-2k} \calK_{\tfrac{d}{2}-\mu-k}^\mathrm{IR}\!\left(\frac{\sigma m^2}{2}\right) \cdot \hat a_k[F] \nonumber \\
&= \frac{m^{d-2\mu}}{(4\pi)^{d/2}\Gamma(\mu)} \sum\limits_{k=0}^\infty \sum\limits_{l=0}^\infty m^{2(l-k)} \frac{(-\sigma/2)^l}{l!} \nonumber \\
&\times\Gamma\left(k-l-\tfrac{d}{2}+\mu\right) \cdot \hat a_k[F]. \label{IRterms}
\end{align}
obviously cannot be resummed this way and contains terms diverging in the massless limit $m^2\to0$.

These results confirm the claims we made at the end of this Letter: the expansion $\hat G_\mathrm{UV} = \hat G_\mathrm{pert}$ \eqref{UVterms} corresponds exactly to all terms coming from the UV region, while the expansion $\hat G_\mathrm{IR}$ \eqref{IRterms} represents additional terms coming from the IR region.

\bibliography{Wachowski2511}

\end{document}